# WEM: A Node Importance Algorithm in Weighted Networks

Linjie Chen [a, 1], Na Zhao [a, 1], Jie Li [b], Zhen Long [a], Ming Jing [a, *], Jian Wang [c, *]

a. *Key Laboratory in Software Engineering of Yunnan Province, School of Software, Yunnan University, Kunming 650091, P. R. China*

b. *Electric Power Research Institute of Yunnan Power Grid, Kunming 650217, P.R. China*

c. *College of Information Engineering and Automation, Kunming University of Science and Technology, Kunming 650504, P. R. China*

## Abstract

In view of the node importance in weighted networks, ***weighted expected method*** (WEM), was proposed in this paper, which take an advantages of uncertain graph algorithm. First, a weight processing method is proposed based on the relationship between the weight of edges and the intensity of contact between nodes, and the calculation method of the contribution of the weight of edges to the node importance is defined, even if there are two converse situations in the reality. Then, because of the use of dynamic programming method, which reduces the complexity of computational time to a linear level, WEM will be more suitable for the calculation in large weighted networks. Owning to its features of calculating, WEM can, to the greatest extent, ensure a precise order of node essential scores for each node. The node connectivity experiment and SIR simulation experiment show that the WEM has higher accuracy and relatively low time complexity.

## Introduction

Network relationships exist within many complex systems in reality, which involve all aspects of human social activities, such as power networks [1, 2], social networks [3, 4], and epidemic spread network [5, 6]. A network is widely used to characterize the complex relationships between different entities[7, 8], and the topological structure characteristics of a network play an important role in network information mining. For a long time, researchers have focused on research of unweighted network, such as the node importance of unweighted network[7], the edge importance of the unweighted network[9], and link prediction based on unweighted networks[10 , 11]. Degree centrality is a classic, simple, and effective node importance algorithm for unweighted networks, which directly measures the number of neighbors of each node. Chen et al.[12] proposed an improved degree centrality algorithm called LocalRank algorithm, which takes into account the information contained in the fourth-order neighbors of each node. ClusterRank algorithm [13] is based on the simultaneous inclusion of the number of neighbors and clustering coefficients of nodes into the calculation, and the algorithm holds that, in the case of the same number of neighbors, a great clustering coefficient of nodes corresponds to their small influence (importance). Kitsak et al.[14] proposed the k-shell centrality and suggested that the topological location of the node contributes more to the importance of nodes than its first-order neighbor[15].

However, weighted networks carry richer information[7] and can better characterize complex systems in the real world. For example, studying the node importance of the traffic network with main roads forming edges, traffic flow forming weights, and road forks forming nodes can help optimize the utilization of traffic resources[16]. Studying social networks with weighted information can help measure and analyze the complex functions and evolution of real society[17]. In studies on the spread of COVID-19, the weighted air transport network is often used as an important research perspective[6], where weight is often used to represent the number of seats on flights between two airports[18]. Conducting research on the topology of large-scale economic networks with trade volume as a weight in the past will help tap nodes with economic potential and improve the stability and robustness of economic systems[19]. Research on the structural characteristics of geographic information will help guide infrastructure construction better[20]. In addition, many cases of using weighted network modeling studies in biological networks[21,22], power networks[23], and social networks[24] exist.

In recent years, the study of the node importance of weighted network has received increasing attention. Antonios Garas et al.[25] proposed the w-core decomposition algorithm, which solved the problem of the proportion of measurement degree to node strength's[18] contribution to node importance through a small number of adjustable parameters. Marius Eidsaa et al.[26] proposed the s-core decomposition algorithm, which also refers to k-core algorithmic ideas and takes into account the strength of nodes, similar to the w-core decomposition algorithm. Although many algorithms can be applied to unweighted and weighted networks, such as the PageRank algorithm derived from web-based ranking networks[27], their disadvantage is that their algorithm parameters depend on empirical decisions. The LeaderRank algorithm[28] solves this problem well and has better convergence and stronger robustness. In a recent study, Gao et al.[29] improved the H-Index algorithm[30] and proposed the HI algorithm that could be used for undirected weighted networks. The betweenness interstitial centrality algorithm (BT)[31], proximity centrality algorithm (CL)[32], eigenvector centrality algorithm (EC)[33], and the HI algorithm described above can evaluate the importance of nodes more accurately, but they all have the disadvantage of high time complexity. The recently developed ASP algorithm proposed by Lv et al.[34] and the improved algorithm based on information entropy [36] proposed by Xue et al.[35] also have this drawback. The evaluation method based on D-S evidence theory[37] combines the degree of nodes and the strength of nodes to consider the importance of nodes, similar to Bayesian probability theory. Liu et al.[38] combined the topology and dynamic characteristics of the network and proposed dynamic sensitivity centrality index to locate the influential nodes. In accordance with the different scales of the network, Zhang et al. [39] designed a multiscale node importance measurement method [39]. These algorithms, which can be applied to weighted networks, have achieved some satisfactory results, but still have some limitations. These algorithms often assume that the connection between nodes is linearly related to the weight of edges, some of which have higher algorithm time complexity and are difficult to apply to large-scale networks. These two aspects still need to be further explored and innovated.

In response to the above questions, this paper proposed the weighted expected method (WEM), which gives an exact order of importance for each node. Experiments on connectivity[40] and SIR propagation models[41] show that the WEM has better performance in accuracy and efficiency. The WEM adopts the method of dynamic programming, which is why the time complexity in the average case is only $O(d_{\max}, m)$, of which $m$ is the total number of edges of the network $G$ and $d_{\max}$ is the maximum degree of nodes in the network.

# Methods

To measure the node importance of weighted networks, the influence of weights needs to be considered on the basis of considering the network topology[42, 43], and the familiar algorithms such as w-core considered the importance of weights and degrees. The proposed WEM adopts a new calculation method: The information that the weight carries and contributes to the importance of the node is depicted as the score of the node under different possibilities, and the final importance score of the node is obtained by combining the importance of nodes in different possibilities.

The WEM is simple and efficient, and it is divided into three steps: (1) According to the relationship between node contact strength and weight, the weight is preprocessed specifically; (2) The weight after processing is regarded as the probability, and the importance of nodes is calculated using dynamic programming algorithm to calculate under different possibilities; (3) The importance score of nodes can be obtained by combining the importance of nodes under different possibilities.

## 2.1. PRE-PROCESSING OF THE WEIGHT VALUE OF THE EDGE

Given a connected weighted network $G(V,E)$, where $V$ is the node set, $E$ is the edge set, the elements $e_{ij}$ in $E$ represent the edges of node $i$ and node $j$, and the weight of the edge $e_{ij}$ is represented by $G$, of which $w_{ij}$ belongs to the real set, the set $W$ stores the weight information of all edges in the network $G$, where the minimum value is $w_{min}$, the maximum value is $w_{max}$, and the standard min-max normalization formula is

$$w'_{ij} = \frac{w_{ij} - w_{min}}{w_{max} - w_{min}} \tag{1}$$

Notably, normalization formula (1) will cause $w'_{ij}$ value to fall on [0, 1] not (0, 1). To avoid $w'_{ij}$ value from being 0 or 1 caused by $w_{ij} = w_{min}$ or $w_{ij} = w_{max}$, the meaning of the minimum edge of the weight value from being ignored after normalization, or magnifying the meaning of the maximum edge of the weight, we improved formula

(1); we added an adaptive parameter $l$ so that $w'_{ij} \in (0,1)$. In addition, the weight of edges has a positive and reverse contribution to the importance of nodes in different weighting networks. For example, in an aviation network, if the weight represents the number of flights, then the great weight represents a closer relationship between nodes[44]. However, in a transportation network, if the weight represents distance, then the connection between the significant representative nodes is weaker. Thus, we designed the following normalization method:

$$\begin{cases} w'_{ij} = \dfrac{w_{ij} - (w_{\min} - l)}{(w_{\max} + d) - (w_{\min} - l)}, & w_{ij} \sim C_{ij} \\ w'_{ij} = 1 - \dfrac{w_{ij} - (w_{\min} - l)}{(w_{\max} + d) - (w_{\min} - l)}, & w_{ij} \sim -C_{ij} \end{cases} \quad (2)$$

where $w'_{ij}$ is the weight of edges $e_{ij}$ after $w_{ij}$ normalization processing; $w_{max}$ and $w_{min}$ are the maximum and minimum values in the weight of network $G$, respectively; $C_{ij}$ is the importance score of node $i$ and node, of which $l$ represents the average weight of edges of network $G$, $l = \dfrac{\sum_{1}^{N_G} w_{ij}}{E_G}$; $E_G$ is the total number of edges of network $G$; and the symbol $\sim$ represents a positive correlation. Formula (2) represents the weight size as a value of 0 to 1; thus, the larger the weight, the closer to 1 the edge is after normalization.

Formula (2) solves the problem of using formula (1) to normalize the result to an integer 0 or 1 so that the distance between the extremum ($w_{max}$ and $w_{min}$) and the boundary matches the value of the average weight after normalization.

We assume that the role that the weight $w'_{ij}$ plays in the next calculation is an independent probability. On this basis, we propose a new weighted node importance algorithm.

## 2.2. DYNAMIC PROGRAMMING ALGORITHM FOR NODE IMPORTANCE CALCULATION

On the basis of normalizing the weight of edges by using the proposed method, we refer to the well-known Possible World Semantics[45] (Pr formula) to redefine the relationship between the weight of the edges and the contribution of the weights to the importance of nodes. Possible World Semantics interprets probability data as a set of deterministic instances called possible worlds, each of which is associated with the probability it observes. We can set the $w'_{ij}$ value corresponding to edge $e_{ij}$ to represent the independent probability[45-47]. Under this assumption, an indeterminate network with $t$ edges would have $2^t$ possible deterministic networks. For weighted networks $G(V,E)$, $G'(V',E')$ is set to represent a network consisting of node sets $V'$ and $E'$, of which $E' \in E$, $V' \in V$, and the probability of $G'(V',E')$ is set in the real world to be $\mathbf{Pr}(G')$. Then,

$$\mathbf{Pr}(G') = \prod_{e \in E'} w'_{ij} \prod_{e \in E/E'} (1 - w'_{ik}) \tag{3}$$

The degree of node $i$ in network $G'$ is set to $deg(i)$, and $c_i$ is set to be a possible importance score, $c_i \in (1,2,...,deg(i))$. Then, the following formula holds:

$$\Pr[deg(i, \ G') \geq c] = \sum_{G \in G_i'' \geq c} \Pr(G') \tag{4}$$

where $G_i''$ is a collection of subnets that belong to all possibilities in network $G'$. Each element of $G_i''$ is made up of node $i$ and its first-order neighbors. The results of formula (2) will be used for subsequent node importance scoring operations.

Formulas (3) and (4) have higher time complexity. Thus, we reduce time complexity to a linear level by using formulas (5), (6), and (7) to implement dynamic programming for calculation of the node importance[45]

$$X(p,q) = w_{ij} X(p-1,q-1) + (1-w_{ij}) X(p-1,q) \tag{5}$$

$p$ and $q$ are available, where $p \in (0,1,2,...,deg(i))$, $q \in (0,1,2,...,p)$, $E(i) = \{e_1, e_2, e_3, ..., e_{deg(i)}\}$ is set to represent the set of all the edges connected to node $i$, subset $E'(i) \in E(i)$ is preset, and $deg(i | E'(i))$ is set to be the degree of node $i$ in subset $G'(V, E \setminus (E(i) \setminus E'(i)))$, where the operation $A \setminus B$ represents the complement of set $B$ in set $A$, and $X(p, q) = Pr[deg(i | \{e_1, e_2, ..., e_p\}) = q]$, that is, the probability with the node degree of $q$ in the presence of only edge $\{e_1, e_2, ..., e_p\}$ for the ordered edge set $\{e_1, e_2, ..., e_{deg(i)}\}$ of node $i$. Now, with integer $q \in [0, p]$ satisfied, we take the value of integer $p$ through $p \in [0, deg(i)]$, a process that can be seen as traversing all the possibilities of the degrees of all subnets in the world of possibilities $G_i''$. The following is the definition of the boundary conditions of $X$ during the calculation process[45]:

$$\begin{cases} X(0,0) = 1; \\ X(p, -1) = 0, \quad \forall p \in [0, \ deg(i)]; \\ X(p, q) = 0, \quad \forall p \in [0, \ deg(i)], \ q \in [p+1, \ q]; \end{cases} \quad (6)$$

Set $\Pr(deg(i) = c) = \sum_{q=1}^{deg(i)} X(deg(i), q)$, and the following formula holds[45]:

$$\Pr(deg(i) \geq c) = \sum_{k=j}^{deg(i)} \Pr(deg(i) = c) = 1 - \sum_{k=0}^{j-1} \Pr(deg(i) = c) \quad (7)$$

The time complexity of the above dynamic programming (the combination of formulas [5], [6], and [7]), of which $d_{max}$ is the maximum degree of node $O(d_{max}, m)$ in network $G$, and $m$ is the total number of edges of the network.

## 2.3. ORDER OF IMPORTANCE OF NODES

After the implementation of the dynamic programming method in the previous section, we obtain the probability distribution of the possible scenarios ($c_i \in \{1, 2, ..., deg(i)\}$) corresponding to the $deg(i)$ importance scores $c_i$ of any node $i$ in network $G$. Furthermore, we obtain the total score corresponding to node $i$ in network $G$:

$$C_i = \sum_{c=1}^{deg(i)} c_i \Pr(deg(i) \geq c) \quad (8)$$

$M$ is set to be a backup of a subset of the node set $V$ during the execution of the WEM to delete nodes, set $V$ is traversed, the steps described above are performed, and all nodes in network $G$ are sorted according to the score of each node. The following is a pseudocode description of the WEM:

---
**WEM**

---

*Iutput*: Weighted network $G(V, E)$

*Output*: $G$ The order of importance of any node $i$ in

1. The weight values in the edge weight set of $G$ are normalized by formula (2), and the results are stored in set $W'$;
2. *for each i when* $i \in V$;
3. Calculate all Pr values of node $i$ according to formulas (5) to (7) and $W'$;
4. Formula (8) is used to calculate $C_i$;
5. *return* A sequence of set $V$ is sorted in reverse order based on the $C_i$ score.

---

Because of the feature of flexible unlimited decimals in its calculating, it ensures a precise essential score for each node, so that the sort will be more precise rather than the sorts of classical WC or K-core. Meanwhile, the calculation of Pr value is realized by dynamic programming, that is why the time complexity of the WEM is $O(d_{max}, m)$. If two or more nodes calculated by this algorithm score the same, then their rankings are considered to be tied. We compared the time complexity of the WEM with that of other algorithms in Table 1, where $n$ is the total number of nodes of the network, and $m$ is the total number of edges of the network.

Table 1: Time complexity comparison table for node importance sorting algorithms in multiple weighted networks

| *WEM* | *BT* | *CC* | *WC* | *EC* | *HI* |
|---|---|---|---|---|---|
| $O(d_{max}m)$ | $O(n^3)$ | $O(n^3)$ | $O(max(e, nlogn))$ | $O(n^2)$ | $O(n^3)$ |

# Experiments and results

Five algorithms were selected for comparison with the WEM, including the classical weighted version of the BT [31], CL [32], EC [33], w-core algorithm [25], and the HI algorithms developed by Lue et al. [30]. To validate the experiment, we chose two mainstream methods to measure the importance of nodes: connectivity test and SIR simulation test.

## 3.1 DATASETS

We verified the accuracy of node importance sorting between the WEM and the compared algorithms on the following eight real networks: (1) Email_dnc: Direct email network in the 2016 Democratic National Committee email leak; (2) Inf_USAir97: American Airlines network; (3) Reptilia_tortoise: The tortoise neural network. (4) Rt_bahrain: Twitter social network; (5) Windsurfers: Natural disaster network; (6) Lesmis: Social network of characters from Victor Hugo's book Les Miserables; (7) Blocks: The symmetrical power of the network from Gordon Royle, University of Western Australia; and (8) C. elegans neural network: The metabolic network of Caenorhabditis elegans. Table 2 lists the basic topology characteristics of the eight real networks. $V$ is the number of nodes, $E$ is the number of edges, $D_{avg}$ is the average degree, $C$ is the global clustering coefficient [48], and $\gamma$ is the homologous coefficient [49].

Table 2. The basic topology characteristics of the eight real networks

| Datasets Discription | $V$ | $E$ | $D_{avg}$ | $C$ | $\gamma$ |
|---|---|---|---|---|---|
| Email_dnc | 1892 | 4466 | 40 | 6.29 | −0.15 |
| Inf_USAir97 | 332 | 2126 | 12 | 0.63 | −0.21 |
| Reptilia_tortoise | 136 | 374 | 19 | 0.87 | −0.35 |
| Rt_bahrain | 4676 | 7979 | 3 | 0.018 | −0.22 |
| Windsurfers | 43 | 336 | 16 | 0.56 | −0.26 |
| Lesmis | 77 | 254 | 6 | 0.57 | −0.17 |
| Blocks | 300 | 584 | 3 | 0.66 | −0.35 |
| C.elegans_neural | 453 | 2025 | 9 | 0.65 | −0.23 |

We take the maximum connected subnet $G$ of the original dataset network above and then perform the operation of each algorithm on the network $G$.

## 3.2 CONNECTIVITY TEST

The connectivity verification method[40] can reflect the advantages and disadvantages of algorithm sorting, mainly reflected in recording the size of the largest connected subnets in the remaining network when deleting nodes in order in the sorting set. If the drop rate is fast, then the network structure collapses rapidly in the process, and the algorithm can efficiently identify important nodes. Otherwise, the algorithm performance is mediocre.

During the execution of the algorithm, the connectivity $r$ is defined as [50]

$$r = \frac{N_v'}{N_v} \quad (9)$$

where $N_v$ is the total number of nodes contained in the largest connected component in the network before the algorithm is executed, and $N_v'$ represents the total number of nodes contained in the current maximum connectivity component during the process of removing nodes from the network.

We use well-known stability and robustness metrics[50-52] to assess the impact on network connectivity:

$$R = \frac{1}{N'} \sum_{n=1}^{N'} r_n \quad (10)$$

where $r_n$ refers to the current ratio of $N_v'$ to $N_v$ in the process of iteratively deleting nodes in the connectivity calculation. $N_v'$ represents the total number of nodes in the original network and $1/N'$ can be used as a normalization factor to ensure that networks of different sizes can be compared. Owing to this, the figure of various dataset for each algorithm may have a vivid show that attached below.

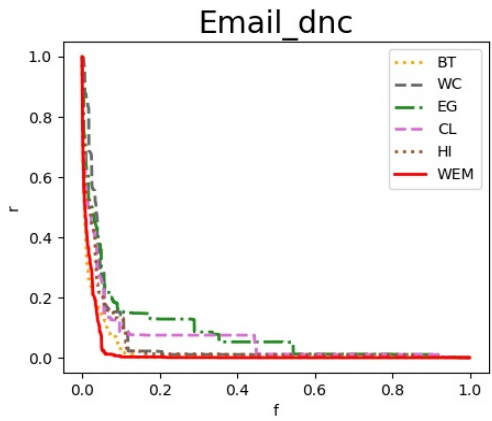

(a)

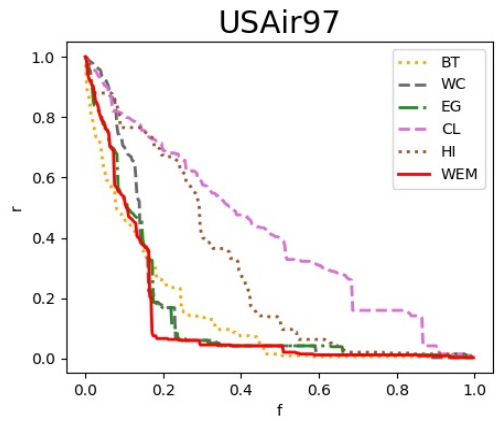

(b)

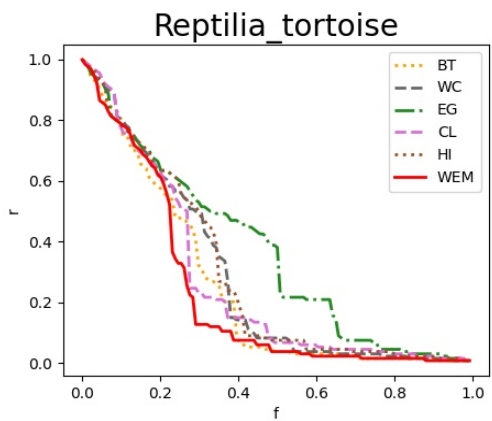

(c)

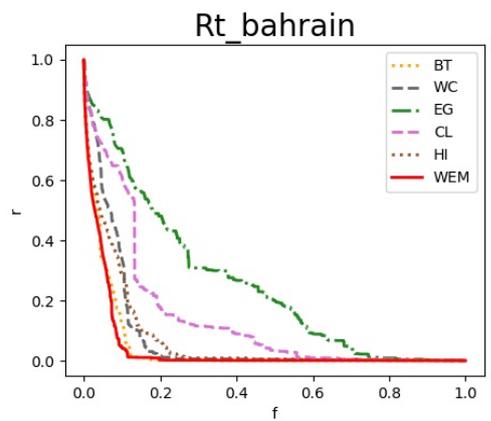

(d)

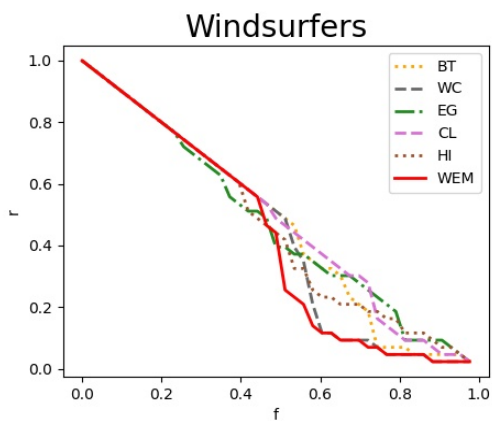

(e)

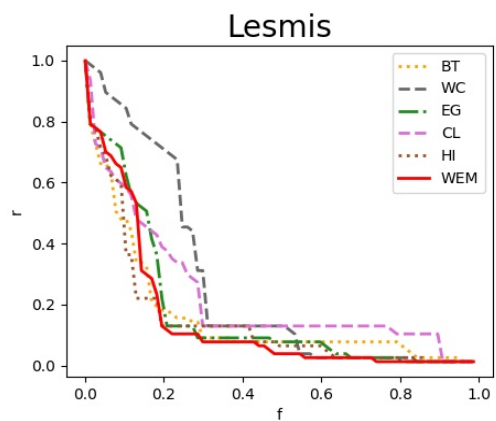

(f)

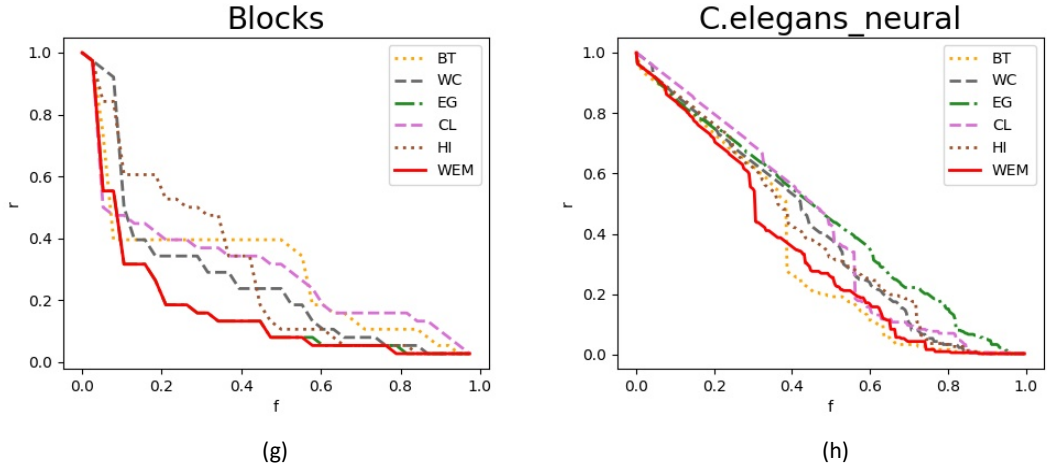

Figure 1: Changes in connectivity in eight real networks when nodes are removed

Table 3: Robustness $R$ values for each algorithm on different datasets

| Robustness:      | WEM   | BT    | EG    | CL    | WC    | HI    |
|------------------|-------|-------|-------|-------|-------|-------|
| Email_dnc        | **0.017** | 0.020 | 0.083 | 0.062 | 0.050 | 0.041 |
| USAir97          | **0.128** | 0.140 | 0.143 | 0.417 | 0.158 | 0.293 |
| Reptilia_tortoise| **0.224** | 0.247 | 0.365 | 0.264 | 0.280 | 0.232 |
| Rt_bahrain       | **0.039** | 0.044 | 0.254 | 0.142 | 0.074 | 0.069 |
| Windsurfers      | **0.426** | 0.476 | 0.479 | 0.493 | 0.444 | 0.468 |
| Lesmis           | **0.151** | 0.164 | 0.177 | 0.232 | 0.269 | 0.152 |
| Blocks           | **0.178** | 0.311 | 0.179 | 0.310 | 0.270 | 0.304 |
| C.elegans_neural | **0.340** | 0.341 | 0.450 | 0.421 | 0.403 | 0.394 |

A small value of $R$ corresponds to faster network collapse and thus, the corresponding sorting algorithm can sort the importance of nodes better. It is not difficult to find that the WEM causes the network to have minimal robustness every time. The connectivity curve of the WEM has a tendency to decline rapidly (which makes the network structure collapse rapidly) in each dataset. We judge the overall robustness of the algorithm by observing the area of the network shape between the connectivity curve and x- and y-axes. As shown in Table 3, the robustness of the WEM on the eight datasets is the lowest among all algorithms as the bold showed, followed by that of the classic BT algorithm. The results indicate that the WEM can make the robustness value the lowest in the process of deleting the nodes of the network $G$, that is, the WEM is more suitable for finding important nodes in the network.

## 3.3 SIR PROPAGATION MODEL TEST

Each time, we take any node $i$ in the target network $G$ as the seed node for infection. In this experimental scenario, 1,000 independent SIR propagation simulations are performed on each and record the average number of recovery nodes at the end of each propagation is taken as the final importance score for the node. Eventually, all the nodes in $G$ are sorted according to their importance score. Then compare the correlations between different algorithms and the Kendall's tau-b correlation of the SIR model sorting results.

There is a need to introduce the Kendall's tau-b correlation coefficients first[53]: $X(x_1, x_2, x_3, ..., x_N)$ and $Y(y_1, y_2, y_3, ..., y_N)$ are respectively the sequence of nodes derived from the evaluation algorithm and the SIR propagation. Integer $s$ and integer $t$ are set, and $s \neq t$, $s \leq N$, $t \leq N$, and the corresponding $s$ and $t$ in sequence $X$ and sequence $Y$ are taken arbitrarily to form two tuples, $(x_s, y_s)$ and $(x_t, y_t)$, respectively. It is a sequence pair if $x_s < x_t, y_s < y_t$ or $x_s > x_t, y_s > y_t$. It is an inverted pair if $x_s < x_t, y_s > y_t$ or $x_s > x_t, y_s < y_t$. Else it is neither a non-sequential pair nor a non-reversed pair. Then Kendall's tau-b is[53]:

$$\tau_b(X,Y) = \frac{2(n_c - n_d)}{N(N-1)} \quad (11)$$

where $n_c$ and $n_d$ are the number of sequential and reverse-order pairs in the $N(N-1)/2$ tuples, respectively.

Notably, in unweighted SIR model, according to mean field theory[54], the prevalence threshold holds:

$$\beta_c \approx \frac{\langle k \rangle}{\langle k^2 \rangle - \langle k \rangle} \quad (12)$$

$\langle k \rangle$ and $\langle k^2 \rangle$ represent the first and second moments of degree $k$, respectively. Mean field theory ensures the expectations of each nod $i$ infecting neighbor nodes $j$ are regular that considered as $Exp(i) = \frac{\sum mp_i}{N}$. $m$ is the average number of neighbors in the target network, and $p_i$ is the probability of infection in the average case ($p_i$ is a constant in the unweighted SIR). WEM is more suitable for weighted networks, to make the experimental results more objective, we selected the weighted SIR(WSIR) [41] for verification, the threshold is:

$$\beta_c ' \approx \frac{\langle k \rangle}{\alpha(\langle k^2 \rangle - \langle k \rangle)} \tag{13}$$

We take $\alpha$ as the average weight of edges in the network $G$, $\alpha = l$ ($l$ has appeared in Methods).

As it is known, each data set has its own particular internal topology, to make it sufficient spread so that acquire an objective result, we multiplied the threshold by a inter 10 before conduct it. The result show in Table 4.

Table 4: Kendall's tau-b correlation coefficients for each algorithm

| WSIR (tau-b):     | WEM   | BT    | EG    | CL    | WC    | HI    |
|-------------------|-------|-------|-------|-------|-------|-------|
| Email             | 0.557 | 0.372 | 0.657 | 0.53  | 0.536 | **0.715** |
| USAir97           | **0.89** | 0.436 | 0.873 | 0.141 | 0.848 | 0.217 |
| Reptilia_tortoise | **0.743** | 0.521 | 0.498 | 0.693 | 0.687 | 0.696 |
| Rt_bahrain        | 0.437 | 0.03  | 0.423 | 0.523 | 0.511 | **0.58** |
| Windsurfers       | 0.743 | 0.375 | 0.322 | 0.322 | 0.615 | **0.745** |
| Lesmis            | **0.831** | 0.272 | 0.685 | 0.274 | 0.796 | 0.674 |
| Blocks            | 0.952 | 0.007 | **0.954** | 0.055 | 0.647 | 0.113 |
| C.elegans_neural  | **0.7** | 0.279 | 0.449 | 0.236 | 0.678 | 0.557 |
| Average value:    | **0.732** | 0.287 | 0.608 | 0.347 | 0.665 | 0.537 |

We can easily see that the WEM in Kendall's tau-b of the SIR model are always the maximum values in most of the dataset and the average value of tau in this compared algorithm is the highest, for the best-performed results are emphasized in bold, indicate that the WEM is closer to the sorting results of a large number of propagation simulation methods in verifying the importance of nodes, that is to say, the important nodes found by the WEM have stronger propagation capability, thereby indicating that the WEM is more suitable for finding important nodes.

# Conclusion and Outlook

Research on node importance in a weighted network is of great significance. Based on the advantages of uncertain graph algorithm, WEM, a node importance mining algorithm(index) based on local topology information was proposed. The weight processing formula is first established according to the correlation between the strength of the connection between nodes and the weight of the edges. Plus, the relationship between node importance calculation and probability calculation of weighted network is defined even if there are two converse situations in the reality. Finally, according to the above process, a simple and efficient node importance algorithm is defined. As experimental verification and comparative analysis show, the WEM has higher sorting accuracy and lower time complexity.

Standing on this basis work, we can look ahead to a challenging topic that can help in the exploration of node importance sorting algorithms based on global information and try to optimize their performance. At present, the mining of relative node importance is a popular topic. In the future, we may try to use the WEM to mine the local topology information of some nodes with known importance in the network to predict the importance of other unknown nodes in the network.